\documentclass[acmsmall]{acmart}

\AtBeginDocument{%
  \providecommand\BibTeX{{%
    \normalfont B\kern-0.5em{\scshape i\kern-0.25em b}\kern-0.8em\TeX}}}

\setcopyright{acmcopyright}
\copyrightyear{2022}
\acmYear{2022}
\acmDOI{}

\acmConference[]{ACM TKDD}{May}{2022}
\acmBooktitle{May '22: ACM Transactions on Knowledge Discovery from Data}
\acmPrice{15.00}
\acmISBN{978-1-4503-XXXX-X/18/06}

\usepackage{balance}
\usepackage{soul}
\usepackage{paralist}
\usepackage{enumitem}
\usepackage{bbm}
\usepackage{float}
\usepackage{subcaption}
\usepackage{wrapfig}

\usepackage{algorithm}
\usepackage{algpseudocode}%

\begin{document}

\title{Supervised Contrastive Learning for Interpretable Long-Form Document Matching}

\author{Akshita Jha}
\affiliation{%
  \institution{Virginia Tech, Arlington, VA}
  \country{USA}}
\email{akshitajha@vt.edu}

\author{Vineeth Rakesh}
\affiliation{%
  \institution{InterDigital, CA}
  \country{USA}}
\email{vineethrakesh@gmail.com}

\author{Jaideep Chandrashekar}
\affiliation{%
  \institution{InterDigital, CA}
  \country{USA}}
\email{jaideep.chandrashekar@interdigital.com}

\author{Adithya Samavedhi}
\affiliation{%
  \institution{Virginia Tech, Arlington, VA}
  \country{USA}}
\email{adithyas@vt.edu}

\author{Chandan K. Reddy}
\affiliation{%
\institution{Virginia Tech, Arlington, VA}
\country{USA}}
\email{reddy@cs.vt.edu}
\renewcommand{\shortauthors}{Jha et al.}


\begin{abstract}

{Recent advancements in deep learning techniques have transformed the area of semantic text matching. However, most state-of-the-art models are designed to operate with {\em short} documents such as tweets, user reviews, comments, etc. These models have fundamental limitations when applied to long-form documents such as scientific papers, legal documents, and patents. When handling such long documents, there are three primary challenges: (i) the presence of different contexts for the same word throughout the document, (ii) small sections of contextually similar text between two documents, but dissimilar text in the remaining parts (this defies the basic understanding of "similarity"), and (iii) the coarse nature of a single global similarity measure which fails to capture the heterogeneity of the document content. In this paper, we describe {\bf CoLDE}: \textbf{Co}ntrastive \textbf{L}ong \textbf{D}ocument \textbf{E}ncoder -- a transformer-based framework that addresses these challenges and allows for interpretable comparisons of long documents. CoLDE uses unique positional embeddings and a multi-headed chunkwise attention layer in conjunction with a {supervised} contrastive learning framework to capture similarity at three different levels: (i) high-level similarity scores between a pair of documents, (ii) similarity scores between different sections within and across documents, and (iii) similarity scores between different {\em chunks} in the same document and across other documents. These fine-grained similarity scores aid in better interpretability. We evaluate CoLDE on three long document datasets namely, ACL Anthology publications, Wikipedia articles, and USPTO patents. Besides outperforming the state-of-the-art methods on the document matching task, CoLDE is also robust to changes in document length and text perturbations and provides interpretable results.} The code for the proposed model is publicly available at \url{https://github.com/InterDigitalInc/CoLDE}.

\end{abstract} 



\maketitle

\section{Introduction}

\begin{figure*}
    \centering
    \includegraphics[width=\textwidth]{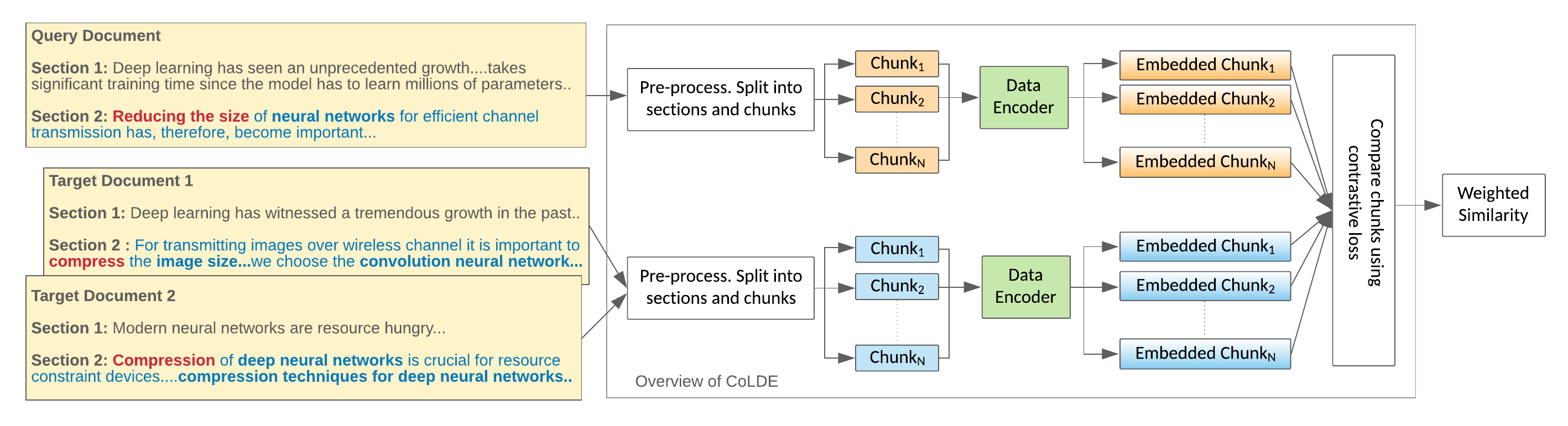}
    \caption{An example illustrating the challenges in long document matching and an overview of the proposed CoLDE framework. The \textit{Query} document is compared against two target documents \textit{Target 1} and \textit{Target 2}. All documents homogeneously discuss \textit{compression} (albeit the context varies). CoLDE provides a weighted similarity score between different chunks and sections within and across the documents.}
    \label{fig:intro}
\end{figure*}

Semantic text matching (STM) is an important and challenging problem in the field of information retrieval. Given a \textit{query}, which could range from a few words to a few pages, the objective of STM is to retrieve a set of documents related to the query. While recent advancements in language modeling have certainly produced promising results, their effectiveness is limited to the following tasks: (a) matching short-queries to short-documents, or (b) matching short-queries to long-documents. A few examples of short-document matching include contextual text similarity \cite{kenter2015short}, sentence matching \cite{hu2014convolutional}, and question ranking \cite{amiri2016learning} where one or both documents (i.e., the query and the target document) have limited textual content. However, \textit{semantic text matching between two long documents}, which is a more challenging problem, has not been thoroughly studied in the literature. Some important use cases for long-document matching include ranking of research papers, comparison of patent documents, and clustering of Wikipedia articles. In this paper, we exclusively focus on \textit{semantic text matching for long documents}. There are three main challenges that arise while developing a model for long document matching. \\

\noindent \textbf{Challenge 1.} It is infeasible to learn the semantics of individual words, phrases, and sentences by considering only a few lines of text, as longer documents span several pages. In other words, learning a representative embedding that encapsulates the context of an entire document is extremely challenging. Further evidence of this is provided by the state-of-the-art Transformer models, such as BERT \cite{devlin2019bert}, which cannot handle more than 512 tokens (or words) during a single feed-forward pass. {Current solutions for addressing these limitations include various pooling techniques which have already proven to be ineffective \cite{reimers2019sentence}}. \\

\noindent \textbf{Challenge 2}. Accounting for various levels of (dis)similarity between different sections of text is another challenge. To understand this better, let us consider a scenario where the end-user is interested in finding long documents most similar to the given document (\textit{Query}). For the sake of simplicity, we consider only two possible target long documents (\textit{Target 1} and \textit{Target 2}) to choose from (see Figure~\ref{fig:intro}). In the above three documents (\textit{Query}, \textit{Target 1}, and \textit{Target 2}), there are three levels of similarity: (a) A similar boiler plate text about the popularity of deep learning that does not contribute to the distinguishing content of the documents, (b) Semantically similar words -- \textit{Query} has a `semantic match' with both \textit{Target 1} and \textit{Target 2} (`reducing the size' is equivalent to the word `compression'), and (c) Contextually similar content -- \textit{Query} is contextually similar only to \textit{Target 2}. The context is provided by statements such as `compressing neural networks' unlike \textit{Target 1} which discusses `image compression'. Here, \textit{Query} is most similar to \textit{Target 2} but in a long-form document, such contextual information cannot be readily learned merely from the initial few lines (or paragraphs) as the distinguishing information might be presented later on in the long-form document.\\

\noindent \textbf{Challenge 3}. The final challenge is interpretability. When comparing documents, it is not sufficient to just provide a score that quantifies the amount of similarity. It is also important to explain what makes the documents (dis)similar.\\

\noindent \textbf{Model Overview.} To tackle the aforementioned challenges, we propose \textbf{CoLDE}: \textbf{Co}ntrastive \textbf{L}ong \textbf{D}ocument \textbf{E}ncoder for matching long documents in an interpretable manner. {Intuitively, while learning document representations, \textit{sections within the same document should be closer to each other and sections across different documents should be farther apart in the latent embedding space}. Additionally, \textit{representations of documents that belong to the same class should be closer to each other compared to the documents that belong to a completely different class.} With this intuition, we leverage the structure of a long-form document along with its class label and build a supervised contrastive learning framework \cite{khosla2020supervised} for long document matching.} CoLDE divides a long document into different sections and uses unique positional embeddings to capture the long document structure for added interpretability. A key challenge in document matching is the {coarse nature of a single global similarity score}. {While some works do demonstrate how attention scores can be used to provide insights into attention-based models \cite{wiegreffe2019attention, yang-etal-2016-hierarchical, ghaeini2018interpreting}}, we define interpretability as the weighted similarity scores between different sections and different chunks of text in a document. These fine-grained similarity scores can aid the end-users in understanding the reason as to why the two documents being (dis)similar. We provide three levels of similarity scores: (i) Similarity score between documents, (ii) Similarity scores between sections within and across different documents, and (iii) Similarity scores between different text \textit{chunks} within and across documents. We use a combination of contrastive loss and a multi-headed attention layer for different text chunks to get this weighted similarity score. Figure~\ref{fig:intro} presents the model overview. We demonstrate the performance of our model using three different semi-structured long document datasets: (i) ACL Anthology Network papers, (ii) Wikipedia articles, and (iii) Patent documents. 

{Recently, a few attempts have been made to encode long sequences of text using transformers \cite{beltagy2020longformer, ding2020cogltx, zaheer2020big}. Our work is significantly different from these approaches since we primarily focus on {\textit{end-to-end long document matching}}. To the best of our knowledge, our work is the first to utilize the long document structure for contrastive learning in text and provide fine-grained analysis of similarity scores within and across documents.}\\
\vspace{-0.05in}

The primary contributions of this paper are as follows:
\begin{itemize}[leftmargin=*]
    \item Propose a {simple yet effective} novel supervised contrastive learning model, CoLDE: Contrastive Long Document Encoder, to compare long documents like research papers, patent documents, and Wikipedia articles.
    \item Develop new data augmentation techniques for text along with unique positional embeddings to capture the long document structure and consequently learn better representations.
    \item Demonstrate the interpretability provided by the CoLDE model -- by providing high-level similarity score between documents, in addition to fine-grained similarity score between different sections, and between different text chunks within and across documents.
\end{itemize}
The rest of this paper is organized as follows. Section 2 describes the related work. Section 3 presents the proposed framework, CoLDE, in detail. Our experimental methodology is discussed in Section 4 and Section 5 concludes our work.

\section{Related Work}

\textbf{Long Document Matching.} Guo \textit{et al.} \cite{guo2016deep} propose a joint deep learning based architecture for ad-hoc retrieval when comparing documents. Several works have also used convolutional networks \cite{hu2014convolutional, pang2016text, yu2018modelling}, with weighting mechanism \cite{yang2016anmm} to generate a final query-document score. Mitra \textit{et al.} \cite{mitra2017learning} propose a combination model that uses local representations and distributed representations for text matching. Yang \textit{et al.} \cite{yang-etal-2016-hierarchical} propose a hierarchical attention network for document classification whereas Adhikari \textit{et al.} \cite{adhikari2019docbert} use BERT for document classifcation. {Jian \textit{et al.} \cite{jiang2019semantic} propose a multi-depth attention based hierarchical recurrent neural network (SMASH) for long-document matching. However, Yang \textit{et al.} \cite{yang2020beyond} pre-train a transformer based hierarchical model (SMITH) for text matching that outperforms them across multiple datasets}. {While our proposed CoLDE model is also based on transformers, we provide fine-grained similarity scores unlike the previous models that only classify whether a document is relevant or irrelevant based on a (coarse) single document level similarity score}.  
 
\noindent \textbf{Long Document Encoding.} Earlier works used Doc2Vec \cite{le2014distributed, chen2017efficient} for long document representation. A growing body of literature is now examining the use of transformer based models for long document encoding \cite{child2019generating, ho2019axial, kitaev2019reformer, liu-lapata-2019-hierarchical, qiu2020blockwise}. Longformer \cite{beltagy2020longformer}, Poolingformer \cite{pmlr-v139-zhang21h}, Performers \cite{choromanski2020rethinking}, and Linear Transformers \cite{katharopoulos2020transformers} adapt transformers to use an attention mechanism to aggregate information for long sequences. Big Bird \cite{zaheer2020big} proposes a sparse attention mechanism that reduces BERT's quadratic dependency on the sequence length to linear. CogLTX \cite{ding2020cogltx} presents a framework that uses multi-step reasoning over key sentences to overcome the insufficient long-range attentions in BERT by using text blocks for rehearsal and decay. Transformer-XL \cite{dai2019transformer} and Compressive Tranformers \cite{rae2019compressive} compress the transformers to use attentive sequence over long text. Our work is significantly different from these approaches since we primarily focus on {\textit{end-to-end long document matching}}.

\noindent \textbf{Interpetability for Text.} Researchers have devised approaches to measure the feature importance in text \cite{ross2017right, sundararajan2017axiomatic}. There is an increased interest in using causal frameworks for understanding black-box predictions \cite{alvarez2017causal}. Some researchers claim that attention is not explanation \cite{jain2019attention}. However, Wiegreffe \textit{et al.} \cite{wiegreffe2019attention},  Yang \textit{et al.} \cite{yang-etal-2016-hierarchical}, and  Ghaeni \textit{et al.} \cite{ghaeini2018interpreting} demonstrate how attention scores can be used to provide insights into attention-based models. We use multi-headed chunkwise attention in conjunction with supervised contrastive loss to provide fine-grained similarity scores that help in interpreting the final decision made by the model.

\noindent \textbf{Contrastive Learning.}
Contrastive learning has seen tremendous success in recent times. SimCLR \cite{chen2020simple} and Supervised SimCLR \cite{khosla2020supervised} use self-supervised and supervised contrastive learning frameworks, respectively, to learn meaningful image representations that help in downstream tasks. DeCLUTR \cite{giorgi2020declutr} uses a self-supervised loss function  for learning universal sentence embeddings without labels. SimCSE \cite{gao2021simcse} uses dropout as a form of data augmentation for learning sentence representations in a self-supervised as well as a supervised manner. Luo \textit{et al.} \cite{luo2021unsupervised} use different semantic paraphrasing techniques for contrastive augmentation to learn better sentence representations. All these methods either focus on images or short-text. To the best of our knowledge, we are the first ones to apply constrastive learning on the task of long-form document matching.

\section{The Proposed C{\small O}LDE Framework}
\begin{figure*}
    \centering
    \includegraphics[width=\textwidth]{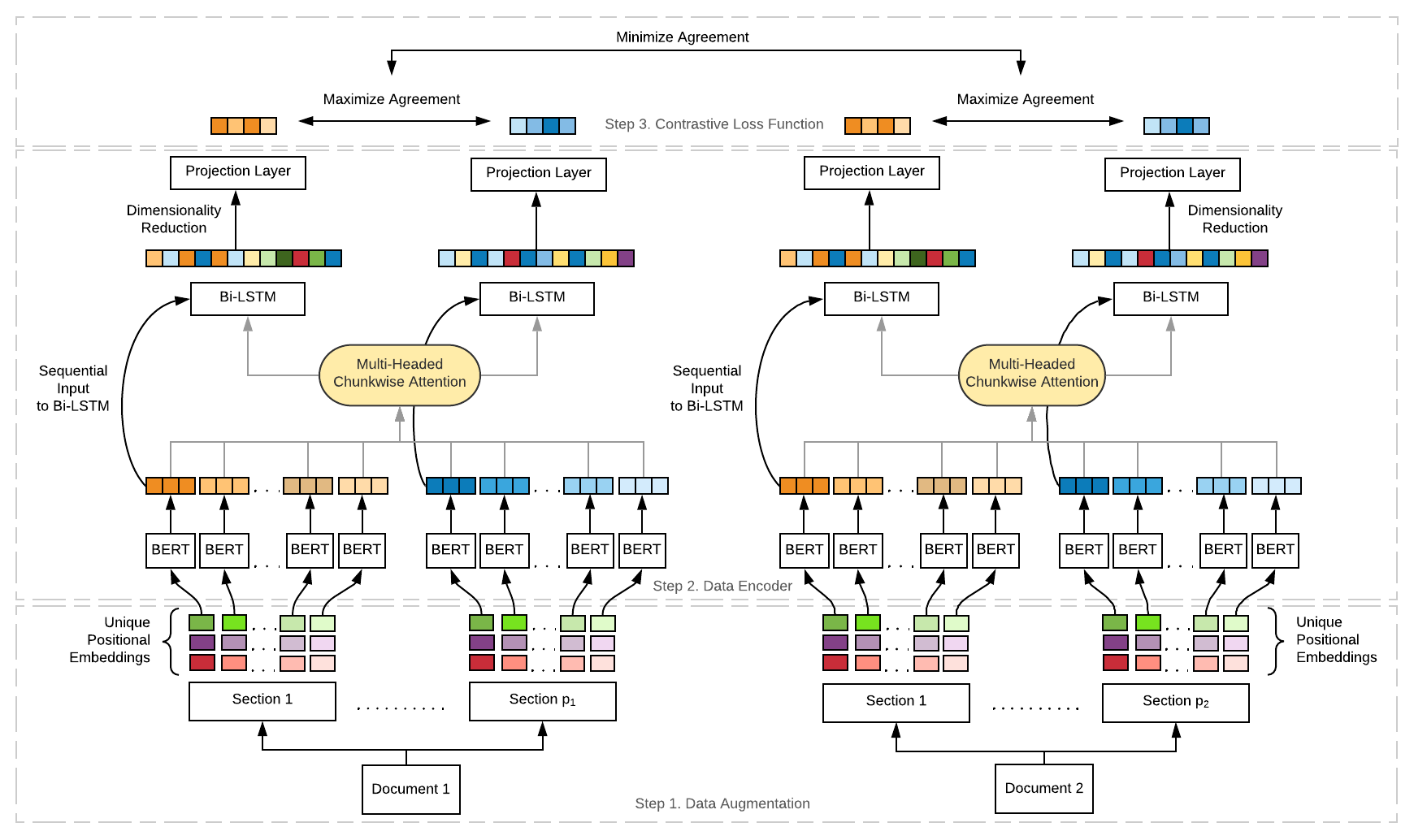}
    \caption{Model architecture for CoLDE. The contrastive learning framework consists of three primary components: (i) Data Augmentation, (ii) Data Encoder, and (iii) Contrastive Loss Function. {It takes as input long documents and divides them into several sections.} Each section is further split into chunks of 512 tokens and enhanced with unique positional embeddings (Fig. ~\ref{fig:uniq_embeddings}) before being given to the data encoder module. The contrastive loss function in conjunction with multi-headed chunkwise attention provide fine-grained similarity scores within and across sections and chunks.}
    \label{fig:full_arch}
\end{figure*}

\begin{figure*}
    \centering
    \includegraphics[width=\textwidth]{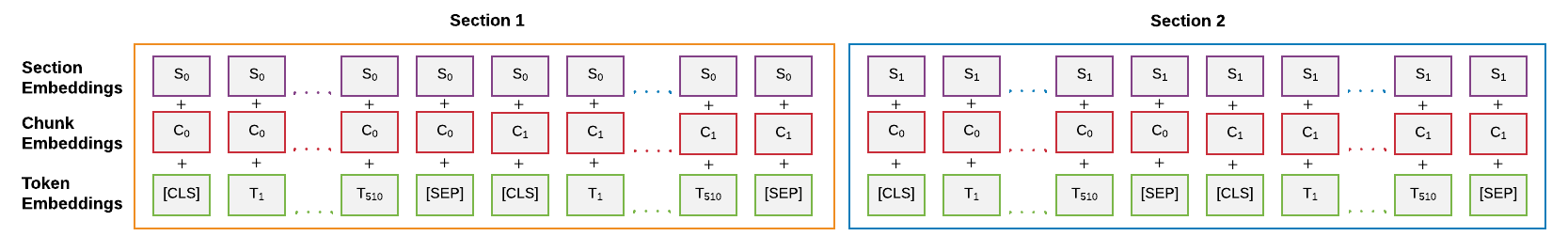}
    \caption{Visualization of the proposed unique positional embeddings for a long-document with two sections. In this example, each section has two chunks of 512 tokens each. The embeddings are summed up together to form the augmented token embeddings for BERT. They capture the structural information present in the long document.}
    \label{fig:uniq_embeddings}
\end{figure*}

Inspired by the recent success of contrastive learning algorithms in the computer vision domain \cite{chen2020simple, khosla2020supervised}, we build a {simple yet effective} contrastive learning framework for long-document matching. Intuitively, CoLDE maximizes the similarity between latent representations of sections within the same document and minimizes the similarity between representations of sections from different documents. Additionally, it ensures that representations of documents belonging to the same class are closer to each other in the latent embedding space  when compared to the documents from a different class. CoLDE consists of three primary components: (i) Data Augmentation, (ii) Data Encoder, and (iii) Contrastive Loss Function. Figure~\ref{fig:full_arch} presents the complete architecture of the proposed model. 

\noindent\textbf{Problem Statement.} {The problem being tackled in this paper is defined as follows. Given a query document $s$, and a set of target documents $D_T$, the goal is to estimate the semantic similarity $sim(.)$ between $s$ and a target document $t \in D_T$ at three different levels: (i) Document level $\hat{y_{d}} = sim(s, t)$ for every document pair, (ii) Section level $\hat{y_{p}} = sim(s_i, t_j)$, where $s_i \in s$ and $t_j \in t$ for every section pair, and (iii) Chunk level $\hat{y_{c}} = sim(c_{q_{s}}, c_{q_{t}})$, where $c_{q_{s}} \in s_i$ and $c_{q_{t}}\in t_j$ for every chunk pair. The semantically similar target documents, sections, and chunks will have a higher similarity score. The notation used throughout the paper and their definitions have been given in Table~\ref{tab:not}.} 

\begin{table}[]
\caption{Table of Notations.}
    \centering
    \small
    \begin{tabular}{|c|l|}
        \hline
        \textbf{Notation} & \textbf{Definitions}\\
        \hline
        $r$ & Token ID \\
        $q$ & Chunk ID \\
        $p$ & Section ID \\
         $x$ & Long document \\
        $s$ & Query document \\
        $t$ & Target document \\
        $f(.)$ & Encoder module \\
        $sim(.)$ & Similarity function \\
        $Aug(.)$ &  Data augmentation module \\
         $\hat{y}_d$ & Document level similarity \\
        $\hat{y}_p$ & Section level similarity \\
        $\hat{y}_c$ & Chunk level similarity \\
        $T_r$ & Embedding for token $r$ \\ 
        $C_p$ & Embedding for chunk $q$ \\ 
        $S_p$ & Embedding for section $p$ \\ 
        \hline
    \end{tabular}
    \label{tab:not}
\end{table}

\subsection{Data Augmentation}
Long documents, unlike short-text, follow an inherent structure and are organized into sections and sub-sections (e.g., introduction, related work, and  methodology). The task of learning meaningful representations for long documents, can be broken down into learning better representations for their sections and sub-sections. To achieve this, we divide a long document into different sections as follows:  $\tilde{x_i}, \tilde{x_j} = Aug(x)$ where $Aug(.)$ is the augmentation module that takes a long document $x$ as input and extracts different sections $\tilde{x_i}$ and $\tilde{x_j}$ from the document. {It should be highlighted that data augmentation is different from data pre-processing as it takes a long document and creates more samples from it.} This is represented as Step 1 in Figure ~\ref{fig:full_arch}.
These sections contain a subset of the text and represent a subset of information present in the original long document. One of the major limitations of BERT is its inability to handle more than 512 tokens at a time, thus making it ineffective for encoding long documents. To overcome this limitation, we divide a document section into multiple chunks of non-overlapping 512 tokens. Additionally, we enhance these input section chunks using unique positional embeddings (described below) to encode the long document structure. 

\subsubsection*{\textbf{Unique Positional Embeddings}}

The token embeddings from BERT are combined with long document structure aware positional embeddings before providing them as input to the BERT model. These positional embeddings are explained below (shown in Fig. ~\ref{fig:uniq_embeddings}).

\begin{itemize}[leftmargin=*]
    \item \textbf{Section Embeddings}: Section embeddings $S_p$ are unique values given to each section where $p$ represents the section number. Since we divide a long document into two sections, these embeddings take two possible values: $0$ and $1$. Embeddings for section $\tilde{x}_i$ take the value of 0 and that for section $\tilde{x}_j$ take the value of 1.
    
    \item \textbf{Chunk Embeddings}: Since BERT can only handle 512 tokens at a time, each section is further divided into several chunks of 512 tokens. Each of these chunks is given a unique ID $C_q$ where $q$ represents the chunk number.
    
    \item \textbf{Token Embeddings}: Token embeddings $T_r$ are the index $r$ of the token in the input sequence. They are the same as the ones seen in the standard BERT model.
\end{itemize}

These additional embeddings are summed up together and form the augmented token embeddings $\tilde{T}_{pqr}$ given by:

\begin{equation*}
    \tilde{T}_{pqr} = \sum_{p=0}^{n_p}\sum_{q=0}^{n_c}\sum_{r=0}^{512} ( S_{p} +  C_{pq} + T_{pqr})
\label{eq:uniq}
\end{equation*} 
where $n_p$ is the total number of sections and $n_c$ is the total number of chunks in a section. There are 512 tokens in each chunk. These augmented token embeddings are given to the data encoder for further processing.

\subsection{Data Encoder}

CoLDE's data encoder $f(.)$ is built over the BERT model \cite{devlin2019bert}. We use a pre-trained BERT model that is fine-tuned during training. The augmented input chunks of 512 tokens are given as input to BERT. Zero-padding is done for chunks smaller than 512 tokens. The encoded BERT chunk embeddings are given to a bidirectional-LSTM (Bi-LSTM) \cite{graves2005framewise} layer for aggregation which uses a unique `multi-headed chunkwise attention' component (described below).

\subsubsection*{\textbf{Multi-headed Chunkwise Attention}}
To get chunk level similarity scores, we introduce a multi-headed chunkwise attention layer. This is a self-attention layer \cite{vaswani2017attention} that computes attention between the BERT embeddings of different chunks of 512 tokens within and across sections. As a consequence, we can obtain a fine-grained understanding of the chunk that attends the most to other text chunks in the document. The chunk with the highest attention weight w.r.t. a \textit{Query} chunk plays the most important part in computing the similarity score. We compute the multi-headed chunkwise attention between sections by treating each chunk in section $p$ as a \textit{Query} $(Q)$. The chunks of section $p+1$ are treated as \textit{Keys} $(K)$ against which the attention is computed. The chunks of section $p+1$ are also used in calculating the \textit{Values} $(V)$. These vectors are of dimension $d$. The weights for calculating the query $Q$, keys $K$, and values $V$ are learned during training. The matrix of the outputs is computed as follows:
$Atention(Q, K, V) = softmax(\frac{QK^T}{\sqrt{d}})V$.

\subsubsection*{\textbf{Dimensionality Reduction}} The encoded BERT output of different chunks, along with the multi-headed chunkwise attention, is given sequentially to a Bi-LSTM layer. The Bi-LSTM layer is required for aggregation of the segmented chunk representations in order to obtain an intermediate high-dimensional section representation $\tilde{z}_l$ corresponding to section $\tilde{x}_l$, where $l$ is the section number. This intermediate section representation $\tilde{z}_l=f(\tilde{x}_l)$ is further given to a projection layer $Proj(.)$ for reducing the dimensionality which results in a final section-level representation $z_l=Proj(\tilde{z}_l)$. The final section representations $z_l$ is used to compute the contrastive loss between sections across query and target documents. The loss is then back-propagated through the network during training. 

\subsection{Contrastive Loss Function}

\subsubsection*{\textbf{Supervised Contrastive Loss}} Using the contrastive learning framework, we bring representations of sections within the same documents closer and obtain a similarity score between sections within and across documents. CoLDE also provides document level similarity scores and ensures that documents belonging to the same class are closer to each other in the latent embedding space. To achieve this, we use supervised contrastive loss \cite{khosla2020supervised} as our objective function. We randomly sample a set of $N$ long documents and their corresponding labels $\{x_k, y_k\}_{k=1}^N$. This results in $2N$ data points since each document is divided into two sections. In this paper, we refer to the set of $N$ documents as a `batch' and the set of $2N$ documents as an `augmented batch'. The supervised contrastive loss function is described as follows:
\begin{equation*}
    \mathcal{L}^{sup} = \sum_{i=1}^{2N}\sum_{j=1}^{2N}\frac{-1}{2N_{y_i}-1}\mathbbm{1}_{i\neq j}\cdot\mathbbm{1}_{\tilde{y_i}=\tilde{y_j}}\cdot log\frac{exp(z_i\cdot z_j/\tau)}{\sum_{k=1}^{2N}\mathbbm{1}_{i\neq k}\cdot exp(z_i \cdot z_k/\tau)}
\label{eq:sup2}
\end{equation*}

In the above equation, $N$ is the batch size; $z_l= Proj(f(\tilde{x}_l))$ where $\tilde{x}_l$ is a section; $f(.)$ is the encoder; and $Proj(.)$ is the projection layer. The three indicator functions: (i) $\mathbbm{1}_{i\neq j} \in \{0, 1\}$ evaluate to 1 iff $i \neq j$; (ii) $\mathbbm{1}_{\tilde{y_i}=\tilde{y_j}} \in \{0, 1\}$ evaluate to 1 when the labels for two sections are the same; and (iii) $\mathbbm{1}_{i\neq k} \in \{0, 1\}$ evaluate to 1 iff $i \neq k$. The symbol $\cdot$ indicates inner (dot) product; $\tau$ denotes a temperature parameter. The final loss is summed across all the samples. {The triplet loss, one of the widely used losses for supervised training, is a special case of the contrastive loss when the batch size $N=2$ and contains only one positive and one negative sample.}

The numerator incorporates all positive sections in an augmented batch, i.e., every section with the same label in the augmented batch is treated as a positive sample. The denominator, on the other hand, performs a summation over the negative samples. To understand the positive and negative samples better, let us consider a simple scenario where $N=3$. The batch has three documents:  $D_1$, $D_2$, and $D_3$ with labels $1$, $1$, and $0$, respectively. Each document is split into two sections which results in an augmented batch of $2N=6$. With respect to the source document $D_1$, sections of $D_1$ and $D_2$ are considered to be positive samples for each other. Sections from $D_3$ are considered to be negative samples. Our loss function encourages the encoder to output section representations for $D_1$ and $D_2$ to be closer to each other in the latent embedding space. Although, the above example only describes the binary class scenario, the loss also generalizes well to a multi-class setting. In such a multi-class scenario, the positive and negative samples are computed w.r.t. to the label of the source document.

\subsubsection*{{\textbf{Negative Samples}}}
For a batch size of $N$, all documents having the same label are treated as positive samples and the others are considered to be negative samples w.r.t. a source document. The negative samples belong to one of the three categories: (i) hard negatives, (ii) semi-hard negatives, and (iii) easy negatives. These three categories represent the distance between the representation of the source document and the negative samples in the latent embedding space, ranging from very close for hard negatives to very far for easy negatives. Most models explicitly require these different kinds of negatives as inputs. This is not the case for CoLDE. When using the supervised contrastive loss, hard negative mining is implicitly performed by the model \cite{khosla2020supervised}. The discriminatory power of the encoder increases with the increase in the number of negative samples w.r.t. a source document, i.e., with increase in batch size, the number of negative samples increase. 

\section{Experiments}
In this section, we empirically evaluate the proposed CoLDE framework by studying the following research questions:
\begin{itemize}[leftmargin=*]
    \item \textbf{RQ1:} Does CoLDE perform better than existing methods for downstream tasks such as long document matching?
    \item \textbf{RQ2:} How does the supervised contrastive loss function and the proposed multi-headed attention layer aid in interpretability?
    \item \textbf{RQ3:} How robust is the model to changes in batch size, document length, and text perturbation?
    \item \textbf{RQ4:} What is the impact of different components on the performance of CoLDE?
\end{itemize}

\subsection{Methodology}

\subsubsection{\textbf{Datasets}} For our experiments, we consider three different long document datasets: (i) ACL Anthology Network Corpus (AAN), (ii) Wikipedia (WIKI), and (iii) Patent (PAT) datasets. 

\begin{itemize}[leftmargin=*]
    \item \textbf{ACL Anthology Network Corpus (AAN)\footnote{http://aan.how/download/\#aanNetworkCorpus}}: The AAN corpus \cite{aan} consists of 23,766 papers written by 18,862 authors in 373 venues related to NLP and forms a citation network. Each paper is represented by a node with directed edges connecting a paper (the parent node) to all its cited papers (children nodes). Papers that have been cited by the parent paper are treated as similar samples. We consider this broad notion of similarity which is in line with the previous works in this area \cite{jiang2019semantic, yang2020beyond}. For every similar sample, an irrelevant paper is randomly chosen to create a balanced dataset. Sets of similar papers are given the same labels. To prevent leakage of information and make the task more difficult, the references and the abstract sections are removed. Papers without any content are also removed. The dataset consists of 12,665 randomly selected research papers split into two sections. On average, each section has 6 chunks. \\
    
    \item \textbf{Wikipedia (WIKI)}\footnote{https://dumps.wikimedia.org/enwiki/latest/enwiki-latest-pages-articles.xml.bz2}: We use the Wikipedia dump, and adopt a methodology similar to the one proposed by Jiang \textit{et al.} \cite{jiang2019semantic} to process this data. We create a dataset of similar Wikipedia articles by assuming that similar articles have similar outgoing links. The Jaccard similarity between the outgoing links of the source and the target articles is calculated. If the Jaccard similarity $> 0.5$, the documents are assumed to be similar, otherwise they are considered dissimilar. {Only articles with two or more similar articles are selected. Sets of similar articles are given the same labels. The dataset consists of 59,911 Wikipedia articles split into two sections. On average, each section has 3 chunks.} \\
   
     \item \textbf{Patent (PAT)}\footnote{https://github.com/google/patents-public-data}: {This dataset consists of 3,000 documents sampled from the publicly available USPTO patents. A patent document is extremely long and primarily consists of (i) Abstract, (ii) Claims, and (iii) Description sections. We make use of the Claims and the Description sections for our experiments. On average, the Claims section has 3 chunks, and the Description section has 27 chunks. Three human annotators were given pairs of documents and were asked to label them as similar or not based on topics of relevance. They referred to the Abstract, Claims, and CPC Codes\footnote{https://www.uspto.gov/web/patents/classification/cpc/html/cpc.html} of the patents to measure the similarity. The final document content similarity label was based on majority vote.}
\end{itemize}

{We report the results of the baseline models on the above datasets.\footnote{Since the pre-processed dataset used in \cite{jiang2019semantic, yang2020beyond, yang-etal-2016-hierarchical} is not public, our results are not exactly aligned with theirs. We use their publicly available code and fine-tune the model on our dataset.}}

\subsubsection{\textbf{Implementation Details}}
The proposed CoLDE model was implemented in PyTorch. For the purpose of our experiments, we use the pre-trained $\text{BERT}_\text{BASE}$ model provided by the Huggingface library\footnote{https://huggingface.co/transformers/model\_doc/bert.html}. It has 12 Transformer blocks, a hidden size of 768 and 12 attention blocks. The BERT model is fine-tuned during training. The Bi-LSTM output dimension is 512. The projection layer reduces the dimension to 256. We use a scaled-dot product attention function \cite{vaswani2017attention} with 4 heads as our multi-headed chunkwise attention function. The learning rate is set to 5e-5. We use an Adam optimizer with a weight decay of 0.01. The value for $\tau$ is fixed to 0.5 for all the experiments. The three datasets are split into 80-10-10 for train, validation, and test sets, respectively. The best performance on the validation set is achieved after training the model for 25 epochs on AAN, after 35 epochs on WIKI, and after 200 epochs on the Patent dataset. We experiment with different batch sizes of 50, 75, 100, and 200. The model training on a batch size of 50 and 75 is done on one 16GB Tesla V100 GPU. For a batch size of 100 and 200, we used one 48GB RTX-8000 GPU.

\subsubsection{\textbf{Baseline Experiments}} We primarily compare our model with the following long document comparison baselines.

\begin{itemize}[leftmargin=*]
    \item \textbf{DSSM \cite{huang2013learning}}: A siamese deep learning model with three fully-connected feed-forward layers with cosine similarity as its objective function. 
    
    \item \textbf{ARC-I \cite{hu2014convolutional}}: A siamese based matching model with three layers of 1D convolutional neural network using cosine similarity.
    
    \item \textbf{Hierarchical Attention Networks (HAN)  \cite{yang-etal-2016-hierarchical} }: 
    A hierarchical structure model consisting of two layers of RNN with attention mechanism using cosine similarity.  
    
    \item \textbf{Siamese-BERT (S-BERT) \cite{devlin2019bert}}: This is a siamese matching model with BERT. To handle long document inputs, S-BERT only uses the first 512 tokens of each document. It uses a pre-trained BERT model which is fine-tuned during training and cosine similarity.
    
    \item \textbf{Siamese Multi-depth Transformed based Hierarchical Encoder (SMITH) \cite{yang2020beyond}}:  
    The current SOTA model for long-form document matching tasks. The model uses a transformer-based hierarchical encoder to obtain sentence-level and document-level representations for long-form documents.
    
    \item \textbf{Siamese Longformer (S-LONG) \cite{beltagy2020longformer}}: A transformer-based model for long sequences with an attention mechanism that scales linearly with sequence length. It takes a maximum input length of 4096 tokens.
\end{itemize}

\subsection{RQ1: Evaluation on Long Document Matching}

We evaluate the performance of our model on the downstream task of long document matching. The document matching problem can be treated as a classification task\footnote{We use classification metrics instead of information-retrieval metrics due to the limitations of the dataset which has very few positive samples (2-3) for every document.}. During evaluation, the model is given a pair of documents and the task is to predict whether the two input documents are similar or not. The training is done using the model architecture described in  Figure~\ref{fig:full_arch}. However, during testing, we do not split the document into sections and use only the Data Encoder module to get a single representation for the query and the target documents, respectively. We then use cosine distance to compute the similarity between these representations. If the document level similarity score is above a certain threshold, the documents are considered to be relevant; otherwise, they are considered to be irrelevant. We report precision (P), recall (R), F1-score (F1), and Accuracy (Acc) for this task in Table \ref{tab:doc_comp}. The utility of the model would depend on the specific use-case, but we observe that CoLDE outperforms all the methods and achieves the best F1-scores and Accuracy on the document matching task on all three datasets. We attribute this superior performance to our new mechanism of exploiting the structure of long documents in conjunction with the contrastive learning framework.

\begin{table*}[h]
    \centering
        \caption{Comparison of different methods on various long-document datasets: (i) AAN, (ii) WIKI, and (iii) PAT. Different metrics such as Precision (P), Recall (R), F1-Score (F1), and Accuracy (Acc) for document matching task are reported. The best result is highlighted in boldface and the second best is underlined. {CoLDE's improvements over other methods are statistically significant with
        $p < 0.05$ as measured by Student’s t-test (whenever CoLDE's performance is the best).} We also present the results for the ablated variants of CoLDE.}
    \footnotesize
    \begin{tabular}{|l|c|c|c|c|c|c|c|c|c|c|c|c|}
        \hline 
        & \multicolumn{4}{c|}{\textbf{AAN}} & \multicolumn{4}{c|}{\textbf{WIKI}} & \multicolumn{4}{c|}{\textbf{PAT}} \\
        \hline
        \textbf{Model} & \textbf{P} & \textbf{R} & \textbf{F1} & \textbf{Acc} & \textbf{P} & \textbf{R} & \textbf{F1}  & \textbf{Acc} & \textbf{P} & \textbf{R} & \textbf{F1} & \textbf{Acc}\\
        \hline
        DSSM & 0.676 & 0.716 & \underline{0.711} & 0.686 & 0.736 & 0.830 & 0.780 & \underline{0.788} & 0.538 & 0.538 & 0.538 & 0.521 \\
        ARC-I & 0.647 & 0.683 & 0.665 & 0.655 & 0.799 & 0.666 & 0.727 & 0.764 & 0.782 & 0.695 & 0.736 & 0.708 \\
        HAN & 0.641 & \underline{0.743} & 0.688 & 0.664 & 0.676 & \underline{0.954} & \underline{0.791} & 0.748 & 0.521 & 0.521 & 0.521 & 0.538 \\
        S-BERT & 0.665 & 0.666 & 0.665 & 0.566 & 0.665 & 0.667 & 0.666 & 0.499 & 0.536 & 0.495 & 0.515 & 0.520\\
        SMITH & \textbf{0.732} & 0.603 & 0.662 & 0.691 & 0.640 & \textbf{0.980} & 0.774 & 0.714 & 0.524 & 0.485 & 0.502 & 0.509 \\
        S-LONG & 0.633 & 0.723 & 0.675 & 0.699 & 0.794 & 0.733 & 0.762 & 0.752 & \textbf{0.848} & 0.533 & 0.655 & 0.521 \\
        \hline
        CoLDE & \underline{0.719} & \textbf{0.756} & \textbf{0.734} & \textbf{0.742} & \underline{0.860} & 0.767 & \textbf{0.809} & \textbf{0.801} & \underline{0.829} & \textbf{0.849} & \textbf{0.839} & \textbf{0.833}\\
        CoLDE$_\text{-Aug}$ & 0.623 & 0.629 & 0.623 & 0.616 & 0.606 & 0.580 & 0.592 & 0.606 & 0.778 & 0.809 & 0.790 & 0.757 \\
        CoLDE$_\text{-LSTM}$ & 0.718 & 0.701 & 0.709 & \underline{0.706} & 0.780 & 0.754 & 0.766 & 0.786 & 0.795 & \underline{0.843} & \underline{0.818} & \underline{0.792} \\
        CoLDE$_\text{-CL}$ & 0.599 & 0.499 & 0.545 & 0.595 & \textbf{0.943} & 0.510 & 0.658 & 0.500 & 0.614 &  0.584 & 0.595 & 0.586\\
        \hline
    \end{tabular}
\label{tab:doc_comp}
\end{table*}

\subsection{RQ2: Case Study on Model Interpretation}\label{sec:interp}

\begin{figure*}[h]
\begin{subfigure}[]{0.3\linewidth}
    \includegraphics[width=\textwidth]{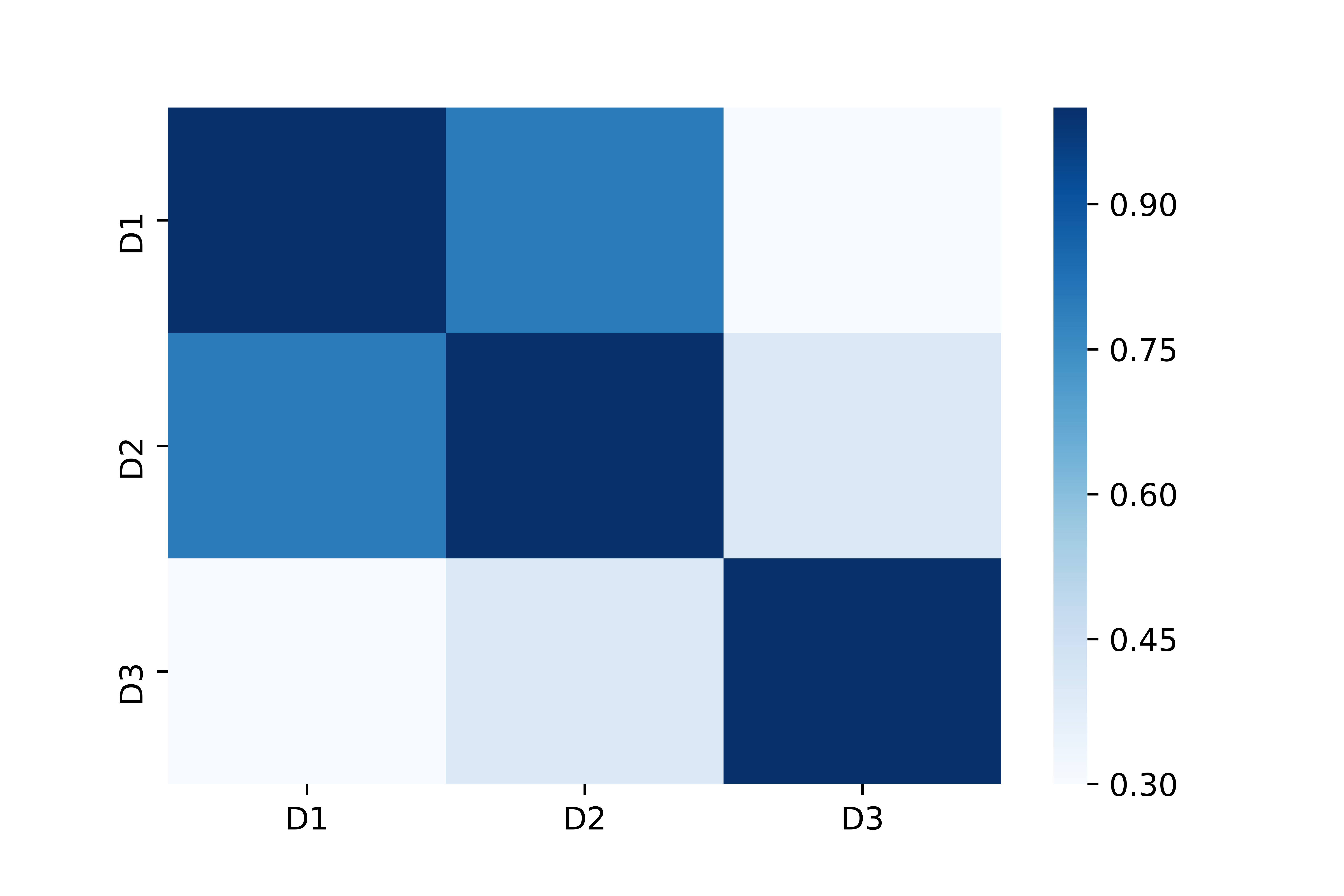}
    \caption{Level 1: Document Level }
    \label{fig:interp_level1}
    \end{subfigure}
    \begin{subfigure}[]{0.3\linewidth}
    \includegraphics[width=\textwidth]{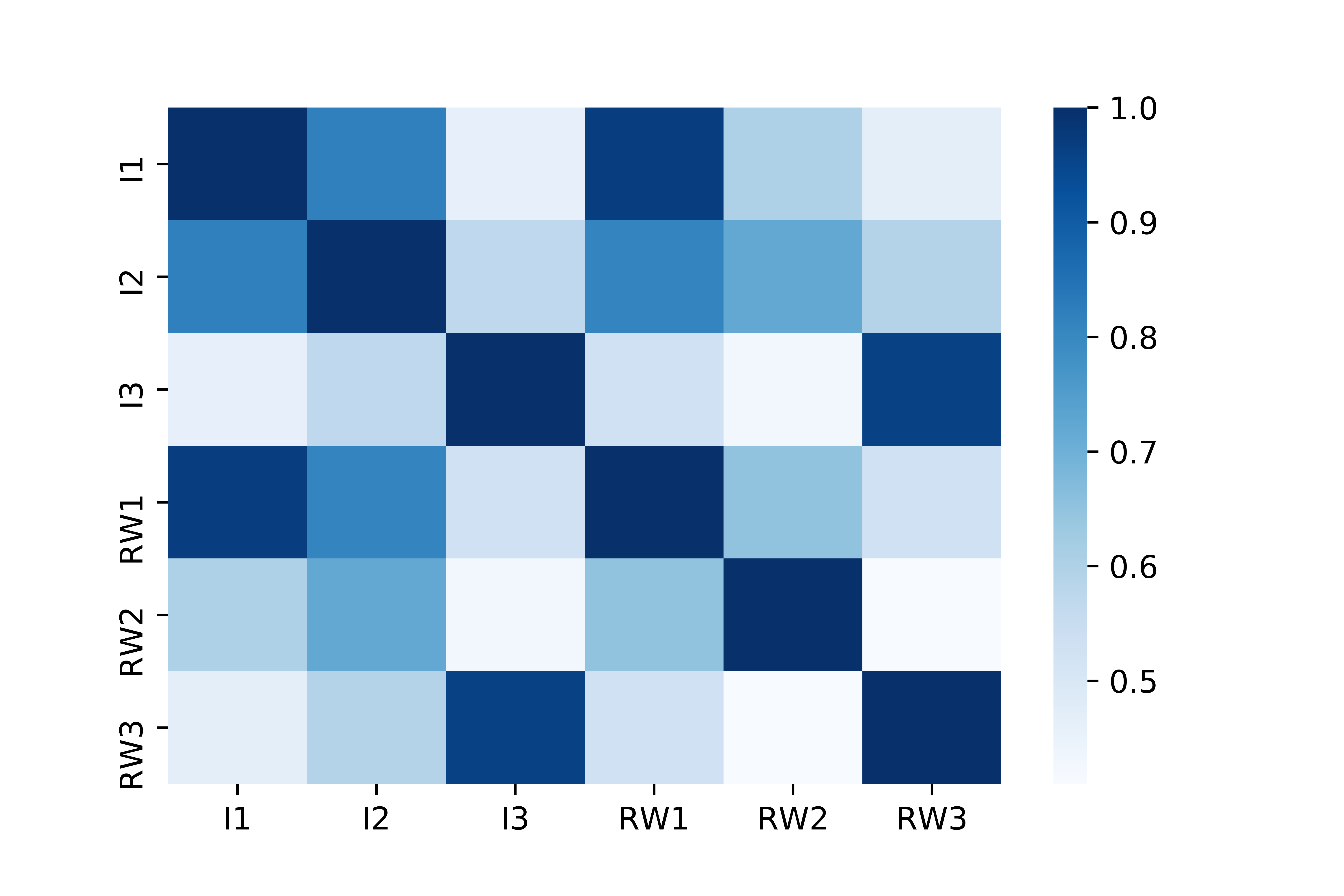}
    \caption{Level 2: Section Level}
    \label{fig:interp_level2}
    \end{subfigure}
    \begin{subfigure}[]{0.3\linewidth}
    \includegraphics[width=\textwidth]{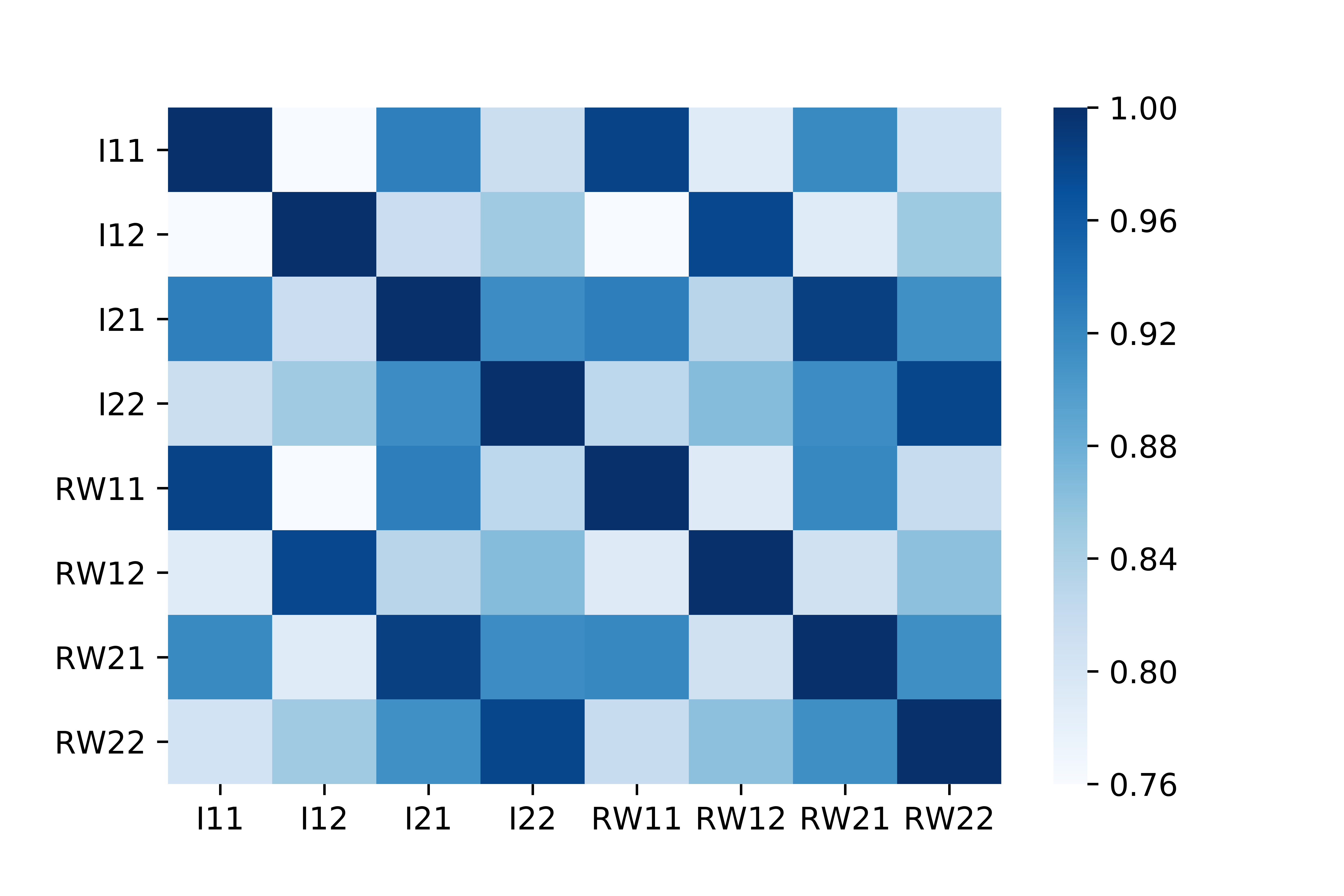}
    \caption{Level 3: Chunk Level}
    \label{fig:interp_level3}
    \end{subfigure}
    \caption{Three levels of interpretability provided by CoLDE: (a) Document Level Similarity, (b) Section Level Similarity, and (c) Chunk Level Similarity. We consider three research papers: $D1$, $D2$, and $D3$. $D1$ and $D2$ are similar to each other. Document $D3$ is different from the other two. $I1-I3$ correspond to the \textit{`Introduction'}; $RW1-RW3$ correspond to the \textit{`Related Work'} sections of documents $D1-D3$, respectively. Qualitative analysis for Fig.~\ref{fig:interp_level3} is shown in Fig.~\ref{fig:qual_interp}.}.
    \label{fig:interp}
\end{figure*}

\begin{figure*}
    \centering
    \includegraphics[width=\textwidth]{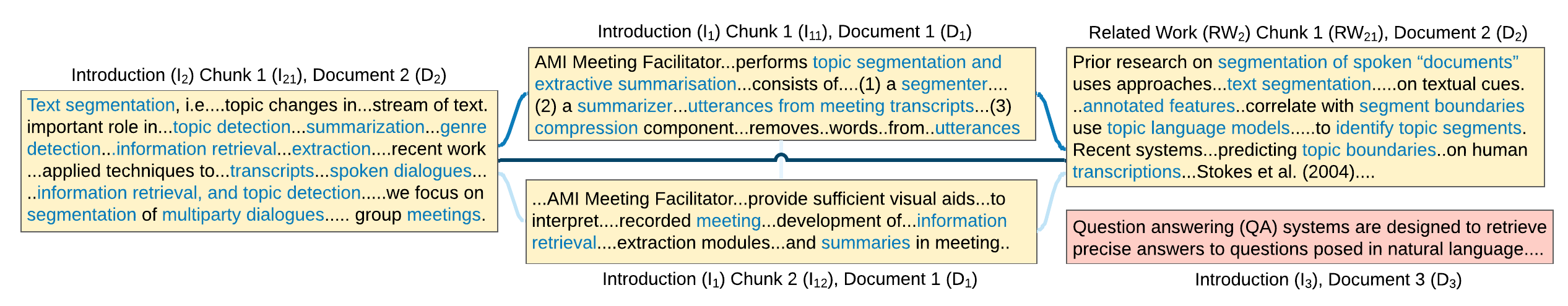}
    \caption{Qualitative Analysis: Documents $D_1$ and $D_2$ are similar. The introduction chunk $I_{11}$ from $D_1$ is about `topic segmentation and extractive summarisation'. The introduction chunk $I_{21}$ from $D_2$ is about `text segmentation'; the related work chunk $RW_{21}$ from $D_2$ presents prior work in the area of `text segmentation'. The introduction chunk $I_{12}$ from $D_1$ summarizes the author contributions. Document $D_3$ is on question answering and is different from the other two documents. Darker colored lines indicate higher similarity.}
    \label{fig:qual_interp}
\end{figure*}

To demonstrate the interpretability provided by CoLDE, we select three documents, $D_1$, $D_2$, and $D_3$ with the following properties. $D_1$ cites $D_2$; hence, they are similar. $D_3$ is not cited by either $D_1$ or $D_2$; hence, $D_3$ is different from other two documents. Figure~\ref{fig:interp} shows the three levels of interpretability (in terms of similarity scores) offered by CoLDE using the architecture describe in Figure~\ref{fig:full_arch}. Qualitative analysis using the text from these documents is shown in Figure~\ref{fig:qual_interp}.

\begin{itemize}[leftmargin=*]
    \item \textbf{Level 1-Document Level Similarity}:
     We obtain the high-level similarity score using CoLDE. Figure~\ref{fig:interp_level1} illustrates the level of similarity between the three documents $D_1$, $D_2$, and $D_3$. A document is most similar to itself. Documents $D_1$ and $D_2$ have a higher similarity score when compared to document $D_3$. Qualitatively, $D_1$ and $D_2$ are about `automatic segmentation of speech' whereas $D_3$ is about `question answering' (see Figure~\ref{fig:qual_interp}).\\

    \item \textbf{Level 2-Section Level Similarity}: While document-level similarity is certainly useful to get a global overview, it does not demonstrate the reason behind the (dis)similarity. Therefore, CoLDE goes a level deeper by providing section level similarity. To explain this outcome, we split each document $\{D_k\}_{1}^{3}$ into two sections $\{(I_k, RW_k)\}_{1}^{3}$, where $I_k$ and $RW_k$ indicate the Introduction and Related Work of $k^{\text{th}}$ document, respectively. Figure~\ref{fig:interp_level2} shows how these are (dis)similar. First, sections from the same document have a high similarity score ($I_1$ and $RW_1$; $I_2$ and $RW_2$; $I_3$ and $RW_3$). Second, since documents $D_1$ and $D_2$ are similar, their introductions $I_1$ and $I_2$ are more similar compared to $I_3$ from $D_3$. This is also true for sections $RW_1$ and $RW_2$, which are more similar to each other compared to $RW_3$. Figure~\ref{fig:qual_interp} shows some of the content from $I_1$, $I_2$, and $RW_2$. The introductions $I_1$ and $I_2$ discuss `topic/text segmentation in dialogues and meetings', and $RW_2$ presents the related work in the area.\\
    
    \item \textbf{Level 3-Chunk Level Similarity}: Besides providing the section-level similarity, CoLDE also explains which segment (or chunk) of a section is important. To illustrate this outcome, we divide $(I_k, RW_k)$ into $j$ chunks, which is denoted by $(I_{jk}, RW_{jk})$. In our experiments, we set $j = 2$; in other words, each document is divided into 4 chunks. Figure ~\ref{fig:interp_level3} shows some interesting traits of chunk-level interpretability between documents $D_1$ and $D_2$. The introduction chunk $I_{11}$ from document $D_1$ is similar to the first introduction chunk $I_{21}$ from $D_2$. In Figure~\ref{fig:qual_interp}, we see that $I_{11}$ broadly discusses `the system components that perform topic segmentation and extractive summarisation', and $I_{21}$ discusses `text segmentation and its applications'. Additionally, the related work chunk $RW_{21}$ from $D_2$ is very similar to both $I_{11}$ and $I_{21}$ from documents $D_1$ and $D_2$, respectively. Qualitatively, $RW_{21}$ presents prior work in the area of `text segmentation'. However, we observe that the second introduction chunk $I_{12}$ from document $D_1$ has very low similarity score with the above chunks ($I_{11}$, $I_{21}$, and $RW_{21}$) and interestingly, $I_{12}$ is about `visual aids for interpretability and a summary of the author contributions'.
\end{itemize}
\noindent {CoLDE, therefore, provides fine-grained \textit{inter-document} and \textit{intra-document} (dis)similarity scores at three different levels which makes the model more interpretable.}

\subsection{RQ3: Robustness Analysis of CoLDE}

\begin{figure*}[h!]
\begin{subfigure}[]{0.3\linewidth}
    \includegraphics[width=\textwidth]{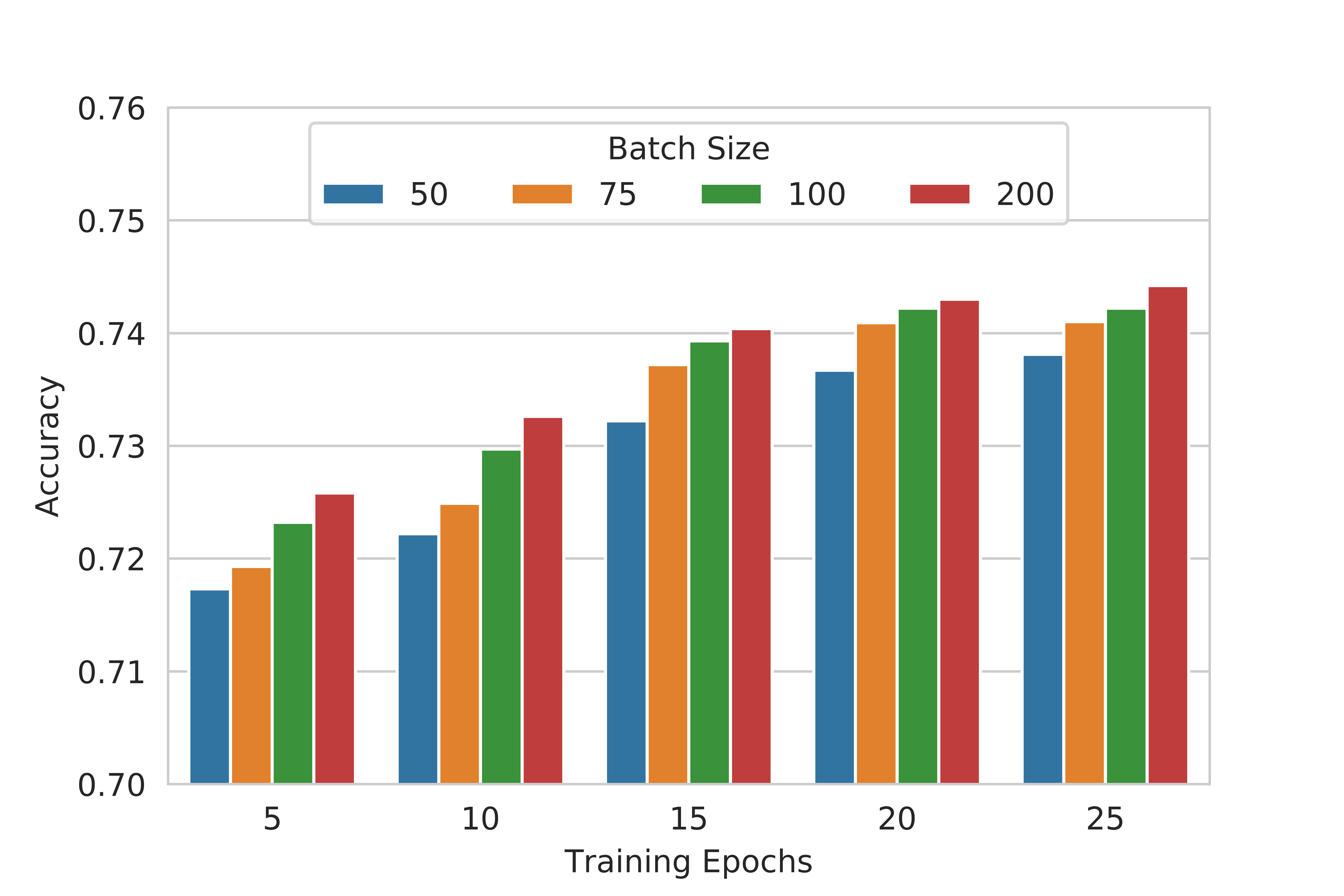}
    \caption{AAN}
    \end{subfigure}
    \begin{subfigure}[]{0.3\linewidth}
    \includegraphics[width=\textwidth]{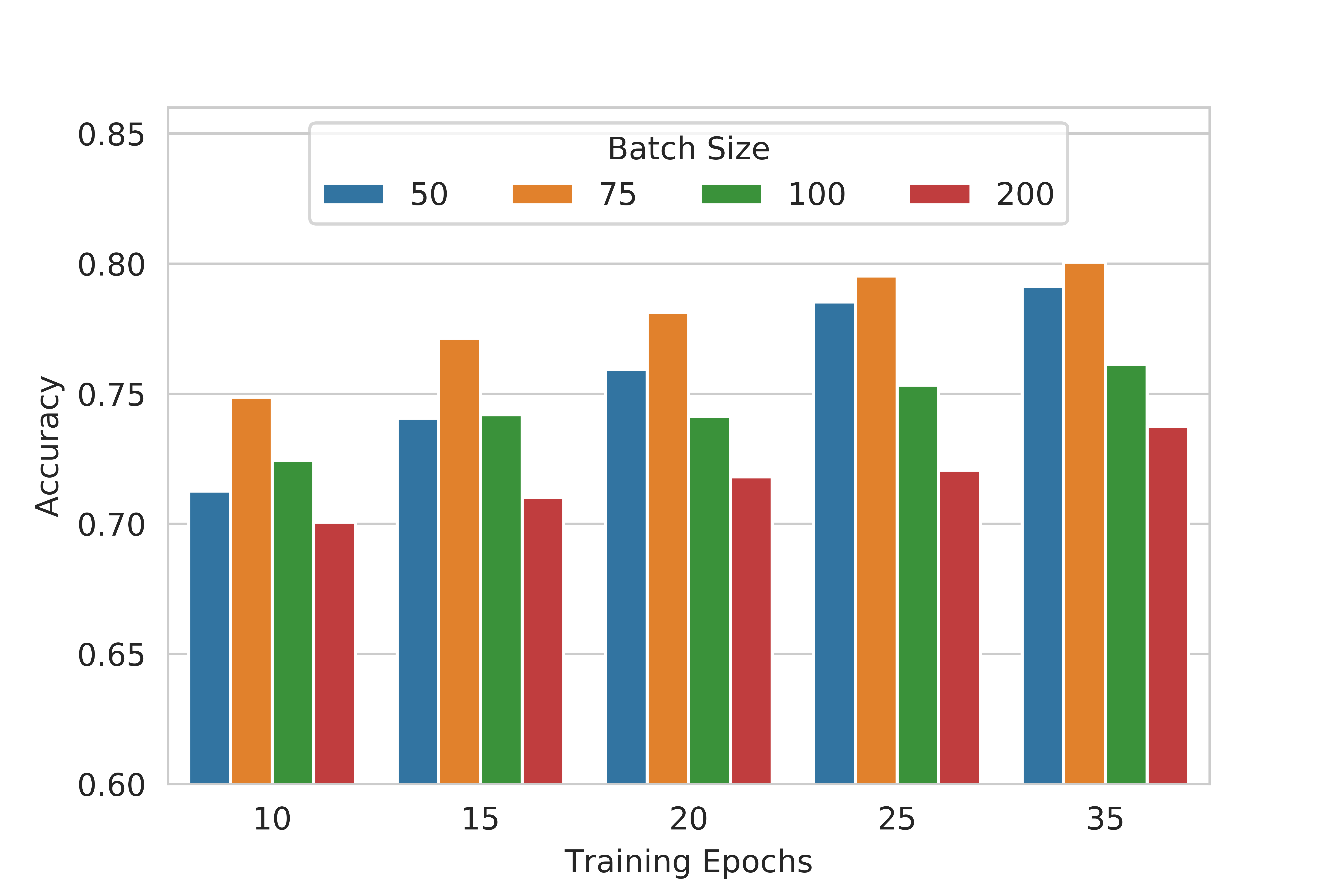}
    \caption{WIKI}
    \end{subfigure}
    \begin{subfigure}[]{0.3\linewidth}
    \includegraphics[width=\textwidth]{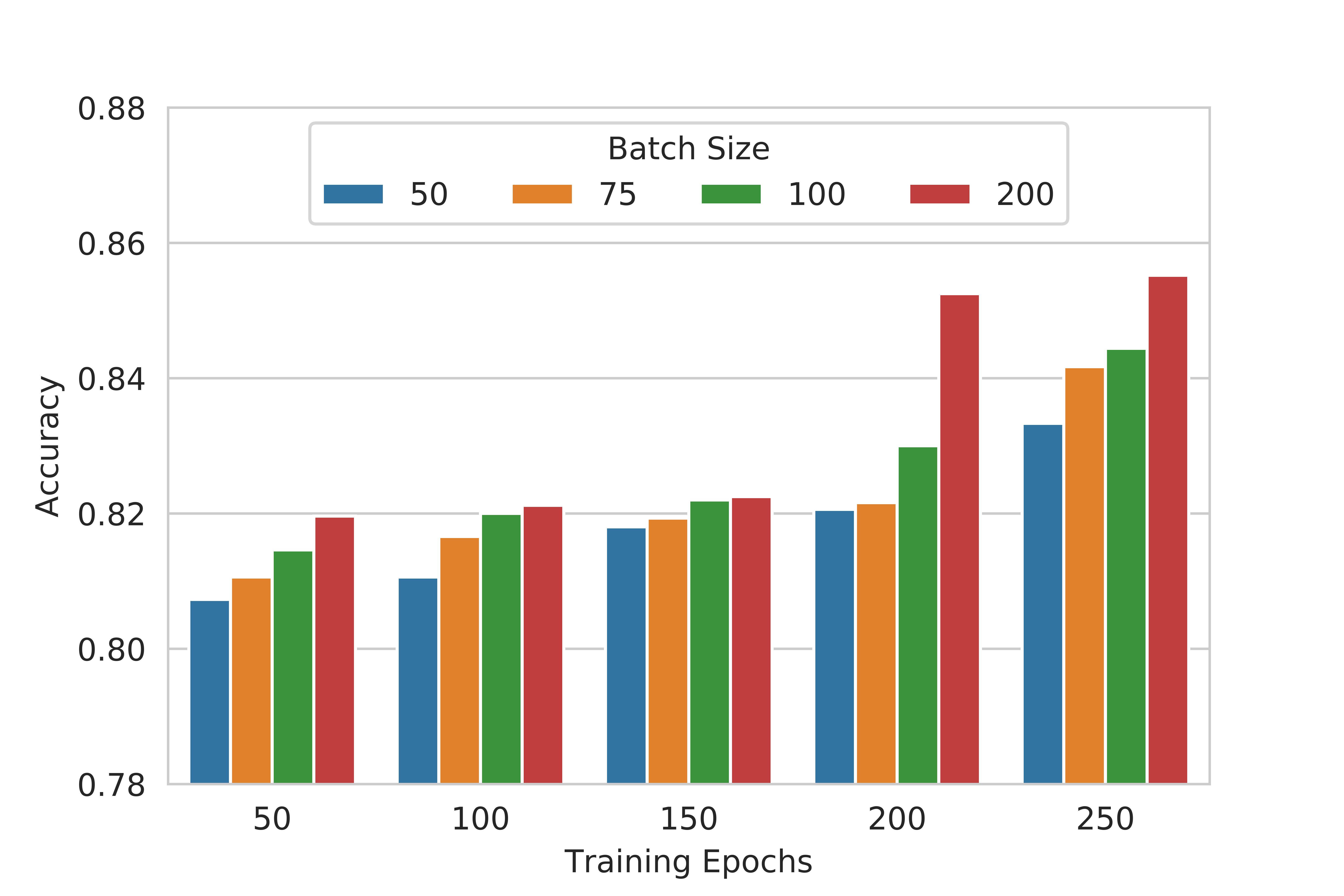}
    \caption{PAT}
    \end{subfigure}
    \caption{Effect of batch size on CoLDE's accuracy for document matching task on the three datasets. }
    \label{fig:acc_batch_size}
\end{figure*}
\vspace{-0.1in}

\subsubsection{\textbf{Effect of Batch Size on CoLDE}}
Recent works on contrastive learning in the computer vision domain \cite{chen2020simple,khosla2020supervised,park2020contrastive} have shown that the discriminating ability of the model increases with increase in batch size since there will be an increase in the number of informative negative samples. We empirically validate this hypothesis and demonstrate that this holds true for the long document matching task. Figure ~\ref{fig:acc_batch_size} shows the impact of batch size on model accuracy when it is trained for the same number of epochs on three datasets: (i) AAN, (ii) WIKI, and (iii) PAT. For the same number of epochs, larger batch-size performs better for AAN and PAT datasets. {However, because of the presence of hard as well as easy negatives in the WIKI dataset, a larger batch size will have many more easy negatives making the task more challenging.} Although, CoLDE does eventually converge to the optimal accuracy of $\sim$80\% for all batch sizes on the WIKI dataset, it takes longer for bigger batch sizes than for smaller ones. Irrespective of the dataset, the best performance of CoLDE steadily increases with the number of training epochs.\\
\vspace{-0.2in}
\begin{figure*}[h!]
    \begin{subfigure}[]{0.3\linewidth}
    \includegraphics[width=\textwidth]{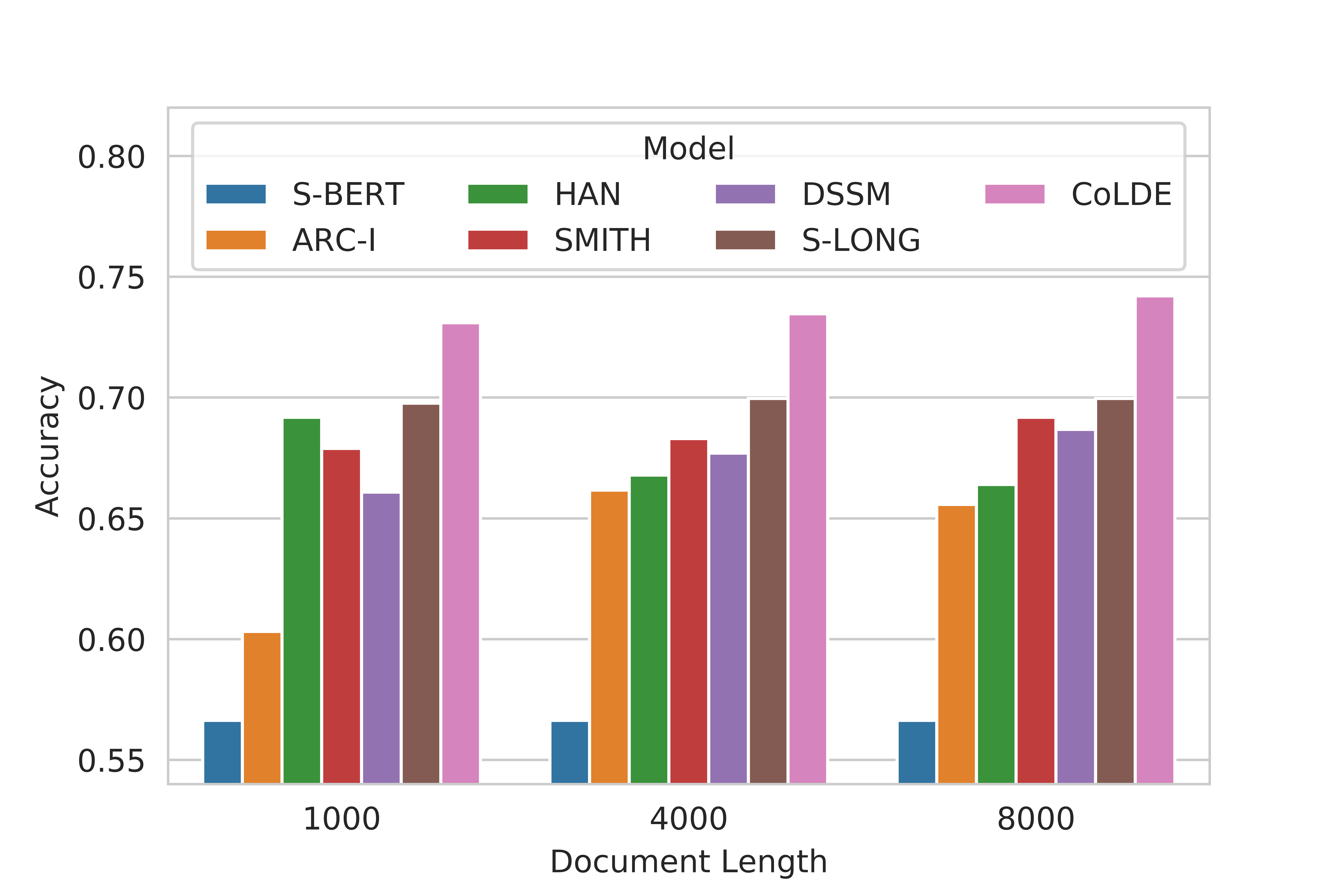}
    \caption{AAN}
    \label{fig:aan_doclen}
    \end{subfigure}
    \begin{subfigure}[]{0.3\linewidth}
    \includegraphics[width=\textwidth]{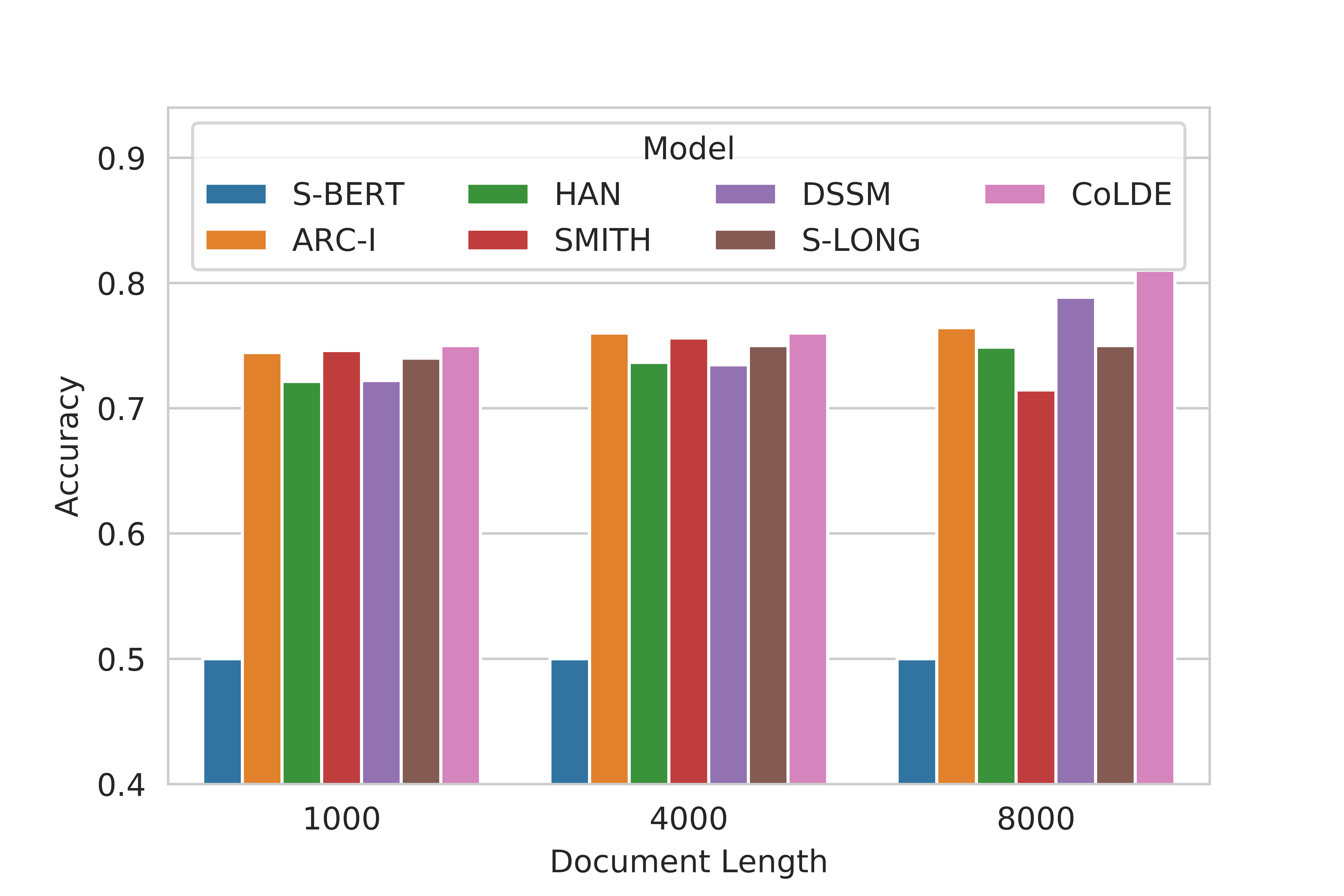}
    \caption{WIKI}
    \label{fig:wiki_doclen}
    \end{subfigure}
    \begin{subfigure}[]{0.3\linewidth}
    \includegraphics[width=\textwidth]{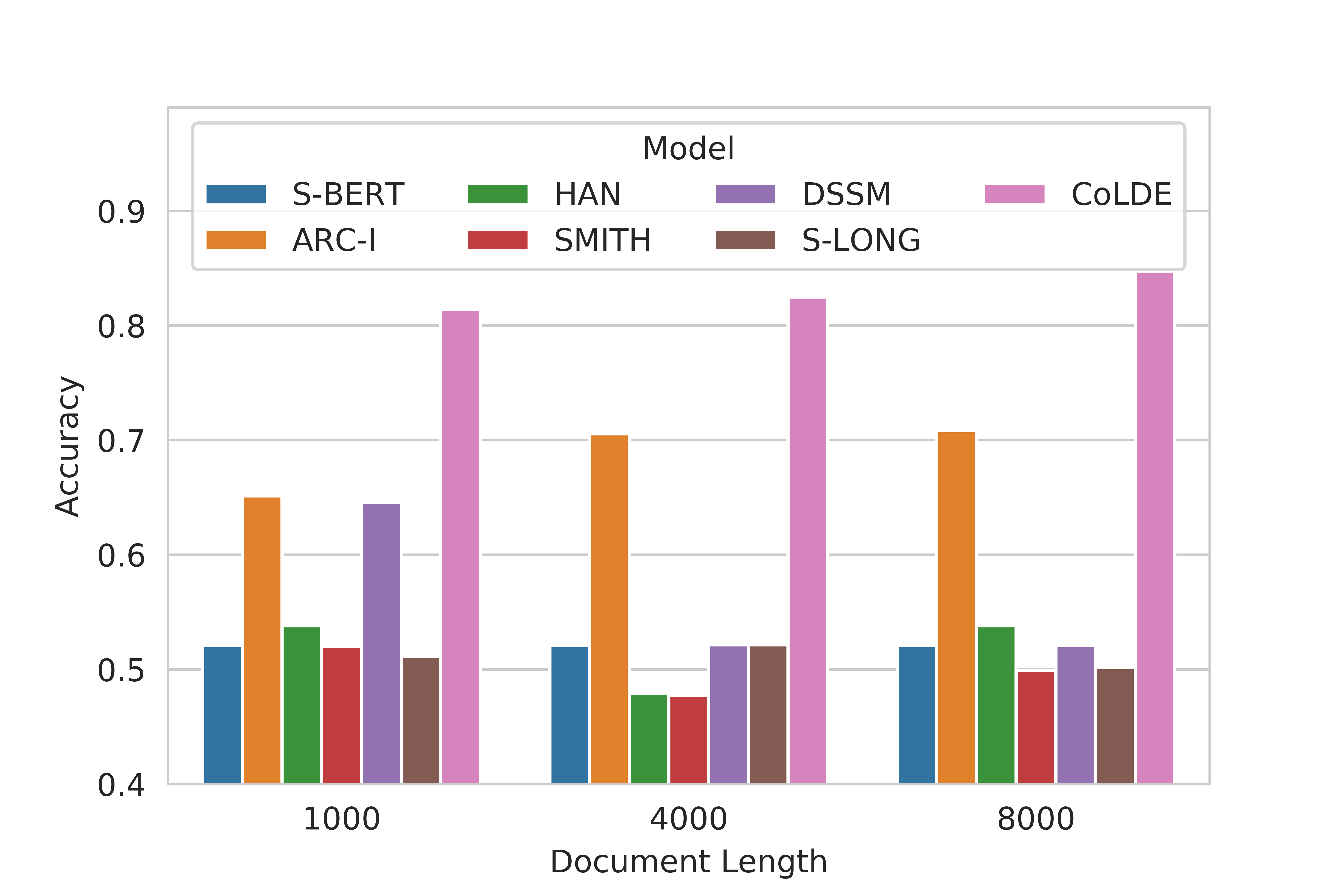}
    \caption{PAT}
    \label{fig:pat_doclen}
    \end{subfigure}
    \caption{Effect of document length on CoLDE's accuracy for document matching task on the three datasets.}
    \label{fig:acc_doclen}
\end{figure*}
\vspace{-0.1in}
\subsubsection{\textbf{Effect of Document Length on CoLDE}} In order to study CoLDE's robustness to varying document lengths, we limit the maximum number of tokens in each section during training and observe the final test accuracy on the document matching task. The model is given the entire long document during evaluation. Figure~\ref{fig:acc_doclen} shows the model accuracy for the maximum lengths of 1000, 4000 and 8000 tokens. We compare CoLDE's performance with other methods for similar document lengths. We observe that CoLDE outperforms the baselines and its performance steadily increases with increase in document length, emphasising the need for explicit long document matching models. \\\vspace{-0.4in}
\begin{figure}[h!]
    \includegraphics[width=0.4\textwidth]{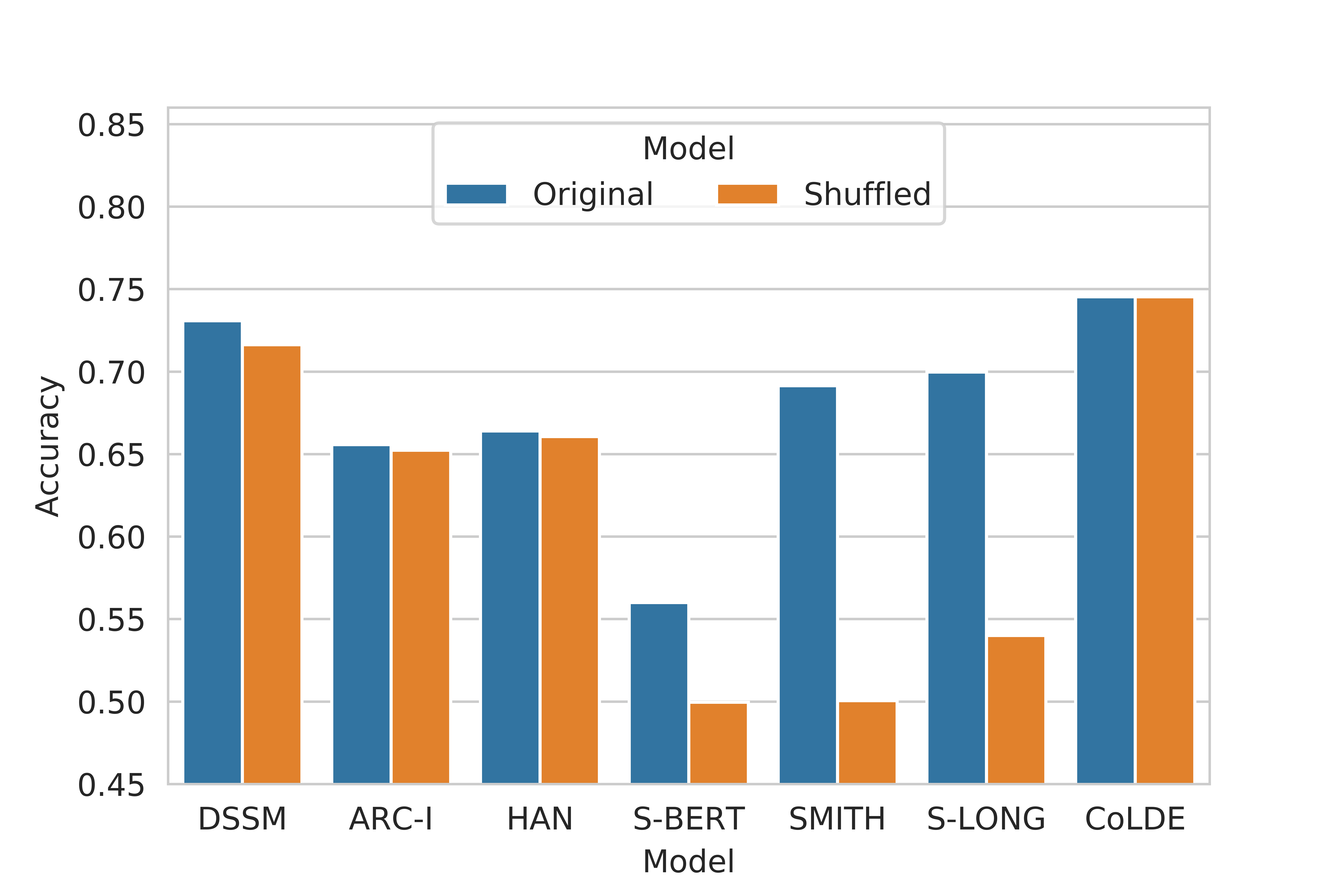}
     \caption{Effect of text perturbation on AAN dataset for document matching task.}
    \label{fig:acc_textperturb}
\end{figure}
\vspace{-0.1in}
\subsubsection{\textbf{Effect of Text Perturbation on CoLDE}} We study the effect of text perturbation on the model performance for the AAN dataset. We randomly shuffle the sections before applying different models and measure their accuracy on the document comparison task. Figure~\ref{fig:acc_textperturb} shows the model performance for all the baseline methods. {We observe a significant drop in the model performance for BERT-based models (S-BERT, S-LONG, and SMITH), whereas CoLDE's performance remains unchanged. There is no significant change in the accuracy for DSSM, ARC-I, and HAN. We observe that, although, other transformer-based models are sensitive to text perturbation, CoLDE is robust. The proposed unique positional embeddings used to enhance the tokens during the data augmentation phase are long document aware, and hence, even after shuffling the text sections, they retain the original long document structure information, thus preserving the accuracy on the downstream task.} 


\subsection{RQ4: Ablation Study of CoLDE}\label{sec:ablation}
We study the effect of different components of CoLDE during training by conducting ablation study experiments. The results of the ablation study on the document matching task are shown in Table ~\ref{tab:doc_comp}. 


\subsubsection{\textbf{Effect of Data Augmentation}} We remove the Data Augmentation module to study its impact on the overall model performance. This model variant is termed as CoLDE$_\text{-Aug}$. We \textit{do not split} a long document into different sections. We also do not add the proposed positional embeddings to the input document. Instead, the entire long document is split into chunks of 512 tokens and given to the data encoder module. The objective function used is supervised contrastive loss. We observe that there is a significant drop in the performance for document matching on all three datasets. The data augmentation module helps in learning better document representations by bringing the representations of sections within a document closer. Moreover, sections of documents belonging to the same class label are also brought closer to each other in the latent embedding space when used in conjunction with the supervised contrastive loss. In its absence, we lose out on learning meaningful document representations, in addition to section level similarity.

\subsubsection{\textbf{Effect of LSTM and Multi-Headed Chunkwise Attention}}
BERT can only handle 512 tokens at a time. To overcome this limitation and get section representations, we use a Bi-LSTM layer for aggregation. We study its effectiveness by removing only the Bi-LSTM and the multi-headed chunkwise attention layer from the data encoder module. We term this variant as CoLDE$_\text{-LSTM}$. The section representation is the `average' of the BERT output representations for all chunks of 512 tokens within a section. This is then given to the projection layer for dimensionality reduction. The objective function used is supervised contrastive loss. We observe that there is a performance drop on the document matching task when not using the Bi-LSTM layer for aggregation. \textit{Bi-LSTM layer effectively captures the sequential nature of the chunks within a section which is lost when computing the average of the BERT output representations for text chunks.}

\subsubsection{\textbf{Effect of using Contrastive Loss}} We conduct an ablation study by replacing the supervised constrastive loss with cosine similarity. Since, we do not require documents to be split into sections for computing the cosine similarity between them, removing the contrastive loss function entails disabling the data augmentation module as well. This variant is termed as CoLDE$_\text{-CL}$. The input documents are split into multiple chunks of 512 tokens and given to the data encoder module. The final representations from this module are used to compute the cosine similarity. From Table~\ref{tab:doc_comp}, we observe that there is a significant drop in performance when using cosine similarity as opposed to using the supervised constrastive loss as our objective function. Supervised contrastive loss helps in providing fine-grained similarity scores and ensures that the latent representations of sections within the same document are closer to each other. Additionally, representations of documents belonging to the same class are also closer in the latent embedding space. 

{\subsubsection{\textbf{Effect of Number of Input Sections}} We study the effect of number of sections on the performance of CoLDE by splitting the source and the target documents into sections ranging from 2 to 5, before being given as input to the model. CoLDE was robust to these changes and the performance was consistent with the ones shown in Table ~\ref{tab:doc_comp} for different input sections for all datasets.}

\section{Conclusion}
In this work, we introduced a contrastive learning framework CoLDE for long-form document matching. CoLDE leverages the long document structure in conjunction with supervised contrastive loss to provide fine-grained similarity scores within and across documents at three different levels: (i) Document Level, (ii) Section Level, and (iii) Chunk Level. The overall framework consists of three primary components: (i) Data Augmentation, (ii) Data Encoder, and (iii) Contrastive Loss Function. We showed that CoLDE outperforms the existing state-of-the-art models for long document matching and presented a detailed case study for model interpretation. We also empirically demonstrated CoLDE's robustness to batch size, document length, and text perturbation compared to other methods. Moreover, we conducted an ablation study to examine the effectiveness of various components of CoLDE. In summary, we demonstrated that CoLDE’s contrastive learning approach is effective for interpretable long-form document matching. 

\begin{acks}
This work was supported in part by the US National Science Foundation grant IIS-1838730.
\end{acks}

\bibliographystyle{ACM-Reference-Format}
\bibliography{sample-base}

\end{document}